\documentclass[11pt]{article}
\usepackage{graphicx}
\usepackage{color}
\usepackage{amssymb}
\usepackage{amsmath}
\usepackage{multirow}
\usepackage{amscd}

\def\Journal#1#2#3#4{{#1} {\bf #2}, #3 #4}

\def\etal{{\it et al.}}

\def\APJ{\em ApJ.}
\def\APJL{\em ApJ.Lett.}
\def\APJS{\em ApJ.Suppl.}

\def\ASS{\em Astrophys.Space.Sci.}

\def\CJA{\em Chin. J. Astron. Astrophys.}

\def\JCA{\em J. Cosmol. Astrop. Phys.}

\def\MRA{\em MNRAS}

\def\PRE{\em Phys. Rep.}

\def\be{\begin{equation}}
\def\ee{\end{equation}}
\def\bea{\begin{eqnarray}}
\def\eea{\end{eqnarray}}

\def\bdir{./} 

\begin{document}
{\Large \bf Simulation of high energy emission from gamma-ray bursts}\\

{\bf Houri Ziaeepour}, {\it Max Planck Institut f\"ur Extraterrestrische Physik (MPE), 
Giessenbachstra$\mathbf{\beta}$e 1, 85748 Garching, Germany.}


\section* {\normalsize Abstract}
Gamma-Ray Bursts (GRBs) are the must violent explosions after the Big-Bang. Their high 
energy radiation can potentially carry information about the most inner part of the 
accretion disk of a collapsing star, ionize the surrounding material in the host galaxy and 
thereby influence the process of star formation specially in the dense environment at high 
redshifts. They can also have a significant contribution in the formation of high energy 
cosmic-rays. Here we present new simulations of GRBs according to a dynamically 
consistent relativistic shock model for the prompt emission, with or without the presence of 
an magnetic field. They show that the properties of observed bursts are well reproduced by 
this model up to GeV energies. They help to better understand GRB phenomenon, and provide an 
insight into characteristics of relativistic jets and particle acceleration which cannot yet 
be simulated with enough precision from first principles.

\section {\normalsize Introduction} \label{sec:intro}
The history of observation of exploding stars goes back quite a long time to 185 
AD~\cite{snhistory}. From this observations we have learned that the life of massive and 
intermediate mass stars - with a mass close or slightly higher than the Sun - ends with 
violent explosions, generally called supernovae. The progenitor of supernovae are divided to 
two main groups~\cite{sntyperev}: Old white dwarfs which arrive to a critical mass - 
Chandrasekhar limit about $1.38 M_\odot$ - by accretion of material from a companion (type Ia), 
and very massive young stars that collapse on themselves and depending on absence or presence 
of hydrogen line in their spectrum, are classified as type Ib/c or type II.

Since 1960 spy satellites called {\it Vela} designed to detect x-ray, gamma-ray, and neutron 
from space and atmospheric nuclear tests observed flashes of gamma-ray of extra-solar system 
origin~\cite{grbdetect,grbhist}. Distribution of their duration shows a clear grouping of 
bursts to short with duration $\lesssim 2$ sec and long with duration $\gtrsim 2$ sec. 
Their occurrence at cosmological distance and their association to supernovae and explosion of 
stars was first suggested in 1986 by B. Paczynski~\cite{cosminsn}. The short bursts are 
believed to have been generated in the collision of compact objects such as two neutron stars 
or a neutron star and a black holes, and long bursts in the core collapse of massive stars.

Motivated by the absence of detection in other wavelengths and by compactness of the source 
(see e.g. ~\cite{piranrev}), a {\it fireball} of strongly interacting $e^\pm$ plasma ejected 
during the explosion has been suggested as the origin of these Gamma-Ray Bursts 
(GRBs)~\cite{cosminsn,fireball1}. In this model the annihilation of $e^\pm$ to photons is 
assumed to be the origin of detected gamma-ray emission. But this model has various problems. 
For instance, it is difficult if not impossible to explain the Fast Rise Exponential Decline 
(FRED) shape of the peaks, their randomness, and long-lasting afterglow which has been 
observed since 1998 for majority of bursts, thanks to angular resolution 
new gamma-ray telescopes such as BATSE, Swift, and Fermi, multi-wavelength detectors on board 
of the Swift and Fermi satellites, and fast slew ability of ground based telescopes. Also it 
cannot explain the power-law spectrum of observed bursts and the lack of a thermal emission 
with a temperature $\sim 1$ MeV.

In the internal shock model, Synchrotron Self-Compton (SSC) emission produced by collisions 
between shells inside a relativistic ejecta are considered to be the origin of observed prompt 
gamma-ray~\cite{intext}. Similarly, the afterglow in lower energies is assumed to be produced 
by the collision of the remnant of the jet with circumburst material or the Inter-Stellar 
Material (ISM). Other models such as a flow of magnetized plasma - a Poynting flow - is 
another popular model for GRBs~\cite{poytingflow}. In this context the gamma-ray is emitted 
by electrons accelerated by reconnection of magnetic field lines. Variants and combination 
of these models are also suggested by various authors to solve some of the short comings of 
these models. 

None of these models is completely flawless. As mentioned above the spectrum of GRBs 
is not consistent with a close to thermal spectrum predicted by a standard fireball model. 
The Poynting flow model cannot explain in a natural way the fast variation of GRB emission 
because the frequency of reconnection is expected to be very low. SSC that is the most favorite 
model of GRB emission has also various issues: To have a sufficiently hard emission the 
magnetic field must be significant such that the emission from most popular electrons with a 
Lorentz factor close to the minimum $\gamma_m$ that make of the peak of spectrum be enough 
hard. This makes the duration of emission of single electrons very short and is known as fast 
cooling problem. Therefore it seems that SSC is not able to sustain long bursts. More 
seriously, synchrotron theory predicts a spectrum index $\alpha \sim -4/3$ at $E \ll 
E_{peak}$~\cite{emission1}, but observations show softer distribution with $\alpha \gtrsim -1$ 
at lower wing is observed.~\cite{batcat}. Recently, observations by the Fermi satellite up to 
energies $\sim 100$ GeV have detected a high energy component in both short and long bursts 
that is delayed by up to few tens of seconds in long bursts from $E \sim 100$ MeV component. 
It fades much slower than lower energies. Finally, SSC has a small efficiency. Particle In 
Cell (PIC) simulations show that only $\lesssim 10\%$ of the total kinetic energy is transfer 
to electrons~\cite{fermiaccspec}. 

At present PIC simulations are not yet able to simulate GRB emission from first principles. 
In this proceedings we review an approximate but realistic formulation of SSC in the context of 
relativistic shocks model~\cite{hourigrb,hourigrb1,hourigrb2}. The aim of this exercise has 
been to see if despite issues discussed above internal shock-SSC model can explain observations.
We also extend the model by considering an external precessing magnetic field to explain 
coherent oscillations observed in GRB 090709A and with less significance in other bursts. 
Then we present light curves and spectra of a number of simulated bursts according to this 
approximation. We show that due to rapid variation of physical quantities, even in presence of 
a precessing field, little evidence of coherent oscillation is imprinted in the emission. 
This explains the lack of observation of a significant oscillatory component in the light 
curves of GRBs. In this proceedings we present a summary of physical processes and motivations 
of the approximations and parameters used in our model, as well as its formulation. Details 
can be found in~\cite{hourigrb,hourigrb2}.

\section {\normalsize Synchrotron emission by relativistic shocks} \label{sec:shock}
In the framework of internal shock model, collisions between shells of material with 
different densities and velocities ejected by a central source produce mildly relativistic 
shocks. They are assumed to be cold and  baryon dominated. Apriori there is no reason 
why faster shells should be ejected later, nonetheless velocity segregation can be 
automatically generated by deceleration of the front shells when they interact with 
surrounding material specially in Wolf-Rayet (WR) stars - the candidate progenitor of long 
GRBs~\cite{shelldecel}. Weak precursors observed in many bursts can be due to this 
process~\cite{precursdecel}.

During a collision compression of the particles behind the shock front and turbulence create 
transverse electric and magnetic fields and produce what is called Electromagnetic Energy 
Structure (EES) - a solitinic electromagnetic wave across the shock front. Particles of 
the slow shell fall along a helical path into the shock front and are accelerated by this 
field and by the short range random fields through Fermi processes. However, their 
penetration distance in the fast shell (upstream) is very short and they are reflected to 
down stream. During this deceleration they emit a fraction of their kinetic energy due to 
the presence of the shock induced magnetic field. The presence of an external magnetic field 
both helps the acceleration of electrons~\cite{fermiaccspec1} and as we see below the 
emission of synchrotron radiation. This process is continuous i.e. electrons move back and 
forth across the shocked zone and in this way, dissipate the kinetic energy of the fast 
shell through synchrotron emission in places where the induced magnetic field is strong and 
transversal. There is phase shift in the EES between electric and magnetic field. Its 
presence is crucial for SSC process in general, and for understanding the origin of high 
energy delayed tail in particular. The lifetime this acceleration-dissipation process is 
short because in a neutral plasma for each electron that falls to the shock front, one or 
more baryons - protons and neutrons - which are more massive fall too. For an observer in 
the rest frame of the slow shell the absorption of these baryons by the fast shell slow it 
down, reduces the discontinuity - the shock - and the strength of EES. We do not consider 
a significant internal energy for the shells and assume that the turbulence and mixing 
transfer the energy from fast shell to slow shell by elastic scattering. In this sense the 
collision is radiative, i.e. all the energy excess is radiated out.

One can distinguish two {\it shocked zone} in the opposite side of the initial discontinuity. 
If the velocity of massive particles - presumably baryons - are reduced to relativistic sound 
speed in the upstream, a secondary {\it reverse shock} front will form which propagates in 
the opposite direction of the main {\it forward shock}. Although it was expected that the 
difference between the dominant synchrotron frequency and time evolution of emission from 
forward and reverse shocks make their separation possible, multi-band and early observations 
of GRBs have shown the contrary. Therefore in the present approximation we only 
consider one radiation emitting region and call it {\it the active region}. Note that what we 
call active region does not correspond to shocked material. In particular, its width 
initially increases to a peak value, then declines and at the end of the collision i.e. when 
the two shells are coalesced, it disappears. This is in contrast of shocked region which 
increases monotonically until shells are completely mixed.

To simplify the model further, we also assume that the thickness of this emitting region is 
small, i.e. the propagation time of photons in this region is smaller than time resolution of 
this model. In fact for objects moving with ultra relativistic speeds with respect to a far 
observer, time and distance are approximately proportional: 
$r' (t') = \beta' (t)ct' \approx ct'$.\footnote{Through this work quantities with a prime are 
measured with respect to the rest frame of the slow shell and without prime with respect to a 
far observer at the redshift of the central engine. Parameters do not have a prime even 
when the parametrization is in the slow shell frame.} Under these approximations evolving 
quantities only depend on the average distance of the active region from central engine. 
Mathematically, this approximation is equivalent to assuming a wavelike behaviour for 
dynamical quantities i.e they depend on $r'- c\beta't'$ rather than $r'$ and $t'$ 
separately. When $\beta' = const$, i.e. when there is no collision or dissipation, this is an 
exact solution. In this case the solution at every point can be obtained from the solution of 
one point. 

In the standard treatment of SSC models~\cite{intext,emission0} a simple power-law 
distribution is assumed for the Lorentz factor of accelerated electrons $n'_e (\gamma_e) = 
N_e ({\gamma_e}/{\gamma_m})^{-(p+1)}$ for $\gamma_e \geqslant \gamma_m$. The parameters used 
for the phenomenological modeling of the shock and SSC emission such as fractions of total 
kinetic energy transferred to accelerated electrons $\epsilon_e$ and to a transversal 
magnetic field $\epsilon_B$ are also considered to be fixed. However, in a phenomenon as fast 
evolving as a GRB these assumptions do not seem realistic. For this reason in our 
formulation it is assumed that $\epsilon_e,~\epsilon_B$, and densities evolve with time. We 
also consider shorter distances for the collision between shells in the range of 
$\sim 10^{10}-10^{12}$ cm rather than $\gtrsim 10^{14}$ cm considered in the literature. This 
leads to short lags between energy bands consistent with observations. The motivation for 
this choice is the detection of variabilities up to the shortest time resolution of present 
instruments - $\sim 10^{-3}$ sec - and the association of anisotropies to the accretion disk 
around the forming compact object, presumably a black hole or neutron star, in the center 
of the collapsing star. 

As for the external magnetic field, it can have various origins: a precessing Poynting flow, 
magnetic field frozen in the plasma, the magnetic field of the central engine or its 
accretion disk, and the dynamo field of the envelop if it is not completely interrupted by 
the explosion. Evidently a combination of all these cases can be present. For this reason and 
also because our simple model cannot distinguish between these field, we do not specify the 
origin of the magnetic field in the mathematical formulation or in the simulations, and 
simply consider a precessing external field i.e. a field with an origin other than the shock.

We have also inspired by findings of PIC simulations and use them as input and/or 
motivation in the choices of parameters and distributions. For instance, simulations of 
relativistic shocks of $e^\pm$ plasma show that the distribution of accelerated electrons is 
close 
to a power-law with exponential cutoff $n'_e (\gamma_e) = N_e ({\gamma_e}/{\gamma_m})^{-(p+1)} 
exp (-\gamma_e / \gamma_{cut})$ or a broken power-law~\cite{fermiaccspec}. We also use 
the penetration distance of accelerated electrons in the slow shell $\lesssim {\mathcal O}(1) 
\times 10^3 \lambda_{ep}$, where $\lambda_{ep}$ is plasma wavelength of electrons, in the 
calculation of Inverse Compton (IC) scattering of photons. This quantity is crucial because 
if we assume the same density in the whole slow shell, most of the photons would be scattered 
before they can leave colliding shells, leading to significant deviation of their spectrum 
from power-law with break expected from synchrotron emission and creates a double peak 
spectrum, in contradiction with observations.

\section{\normalsize Formulation}
Conservation of energy and momentum determines the evolution of the shock. The velocity 
$\beta'$ of the fast shell/active region decreases due to absorption of particles from slow 
shell and dissipation of kinetic energy as radiation due to synchrotron and self-Compton 
interactions. After a variable change the dynamic equations - energy-momentum conservation 
equations - for the active region can be written as:
\bea
&& \hspace{-0.7cm}\frac {d(r'^2 n' \Delta r' \gamma')}{dr'} = \gamma' \biggl (r'^2 
\frac{d(n'\Delta r')}{dr'} + 2r' (n'\Delta r')\biggr ) + 
r'^2 (n'\Delta r') \frac{d\gamma'}{dr'} = n'_0(r) r'^2 - \frac{dE'_{sy}}{4\pi m c^2dr'} 
\label {enercons} \\
&& \hspace{-0.7cm}\frac {d(r'^2 n' \Delta r' \gamma' \beta')}{dr'} = \beta' \gamma' (r'^2 
\frac{d(n'\Delta r')}{dr'} + 2r' (n'\Delta r')) + r'^2 (n'\Delta r') 
\frac {d(\beta' \gamma')}{dr'} = - \frac{dE'_{sy}}{4\pi m c^2dr'} \label {momcons}
\eea
where $r'$ is the average distance of the active region from central engine, $n'$ is the 
baryon number density of the fast shell measured in the slow shell frame, $n'_0$ is the 
baryon number density of the slow shell in its rest frame and in general it depends on $r'$. 
Here we assume that $n'_0(r') = N'_0 (r'/r'_0)^{-\kappa}$. For the ISM or thin shells where 
density difference across them is negligible $\kappa = 0$, i.e. no radial dependence. For a 
wind surrounding the central engine $\kappa = 2$. If we neglect the transverse expansion 
of the jet, for a thin shell/jet expanding adiabatically $\kappa = 2$ too. If the lifetime 
of the collision is short we can neglect the density change due to expansion during the 
collision and assume $\kappa = 0$. $\Delta r'$ is the thickness of the active region, 
$\gamma'$ is the Lorentz factor of the fast shell with respect to the slow shell, 
$\beta' = \sqrt {\gamma'^2 - 1} / \gamma'$, $m = m_p+m_e \approx m_p$, $E'_{sy}$ is the total 
emitted energy, and $c$ is the speed of light. The evolution of the average radius of the 
shell is:
\be
r' (t') - r' (t'_0) = c \int_{t'_0}^{t'} \beta' (t'') dt'' \label {revol}
\ee
where the initial time $t'_0$ is considered to be the beginning of the collision. Equations 
(\ref{enercons}) and (\ref{momcons}) can be solved exactly for the column density of the 
active region $n'(r')\Delta (r')$, but $\beta' (r')$ does not have an analytical solution. We 
use an iterative technique which allow to determine the solution as a function of the 
{\it coupling} defined as: 
\be
{\mathcal A} \equiv \frac{4\alpha m_p^2 \sigma_T n'_0 \Delta r' (r'_0) \epsilon_e^2 (r'_0) 
\epsilon_B (r'_0)}{3 m_e^2} \label{biga}
\ee
In presence of an external magnetic field a second interaction term is also present which its 
{\it coupling} can be defined as:
\be
{\mathcal A}_1 \equiv \frac{\alpha m_p \sigma_T \Delta r' (r'_0) \epsilon_e^2 (r'_0) 
{B'}_{ex\bot}^2 (r'_0)}{6\pi c^2 m_e^2} \label{bigap}
\ee
Details of this calculation is described in ~\cite{hourigrb,hourigrb2}

For the determination of synchrotron flux we use textbook formulations. Nonetheless, we have 
to integrate over angular distribution of emission for the observer. Notably, we must take 
into account the fact that due to relativistic effects even the emission from a spherical 
shell seems highly collimated at far distances. The angle of collimation is 
$\sim 1/2\Gamma (r)$ where $\Gamma (r)$ is the Lorentz factor of the active region. 
The details of this calculation is discussed~\cite{hourigrb}. Finally, after integration 
over emission angle the expression for the received synchrotron flux is:
\be
\frac {dP}{\omega d\omega} = \frac{4\sqrt {3} e^2}{3\pi} r^2 \frac{\Delta r}{\Gamma (r)} 
\int_{\gamma_m}^\infty d\gamma_e n'_e (\gamma_e)\gamma_e^{-2} K_{2/3} (\frac{\omega'}
{\omega'_c}) + {\mathcal F}(\omega, r) \label{powerdopcorr}
\ee
where ${\mathcal F}(\omega, r)$ includes subdominant terms and terms depending on the 
curvature of the emission surface. In~\cite{hourigrb} it is argued that these terms are 
much smaller than the dominant term, thus we neglect them in the simulations.

The advantage of the approximation presented here is that one can use the approximate 
analytical solutions to study the effect of various parameters and quantities on the 
evolution of dynamics of the ejecta and its synchrotron emission. However, the price to pay 
for this simplification is that we cannot determine the evolution of $\Delta r'(r')$ from 
first principles and must consider a phenomenological model for it. In our simulations we 
have used following phenomenological expressions:
\bea
&& \Delta r' = \Delta r'_{\infty} \bigg [1-\biggl (\frac{r'}{r'_0} \biggr )^
{-\delta}\biggr ] \Theta (r'-r'_0) \quad \text {Steady state model} \label {drquasi} \\
&& \Delta r' = \Delta r_{\infty} \bigg [1- \exp (-\delta' \frac{r'}{r'_0}) \biggr ] 
\Theta (r-r'_0) \quad \text {Exponential model} \label {expon} \\
&& \Delta r' = \Delta r'_0 \biggl (\frac {\gamma'_0 \beta'}{\beta'_0 \gamma'} 
\biggr )^{\tau}\Theta (r'-r'_0) \quad \text {dynamical model} \label {drdyn} \\
&& \Delta r' = \Delta r'_0 \biggl (\frac{r'}{r'_0} \biggr )^{-\delta} 
\Theta (r'-r'_0) \quad \text {Power-law model} \label {drquasiend} \\
&& \Delta r' = \Delta r'_0 \exp \biggl (-\delta'\frac{r'}{r'_0} \biggr )
\Theta (r'-r'_0) \quad \text {Exponential decay model} \label {expodecay}
\eea
The initial width $\Delta r'(r'_0)$ in the first two models is zero, therefore they are 
suitable for the description of formation of an active region at the beginning of internal 
or external shocks. The last three models are suitable for describing more moderate growth 
or decline of the active region.

\section{Simulations} \label{sec:simul}
Table \ref{tab:param} shows the list of parameters of this model. Due to their large number 
it is not possible to explore the totality of parameter space. Therefore, in this section we 
show a few examples of simulated light curves and spectra of GRBs. Each simulation consists 
of at least 3 time-intervals (regimes) during which exponents are kept constant. Moreover, 
each time interval corresponds to a given model for the evolution of the width of the 
active region. The first regime must be either steady state or exponential for in which the 
initial width of the active region is zero. Following regimes can be either dynamical or  
power-law. Matching between values of evolving quantities at the boundary of regimes assures 
the continuity of physical quantities.

\begin{figure*}
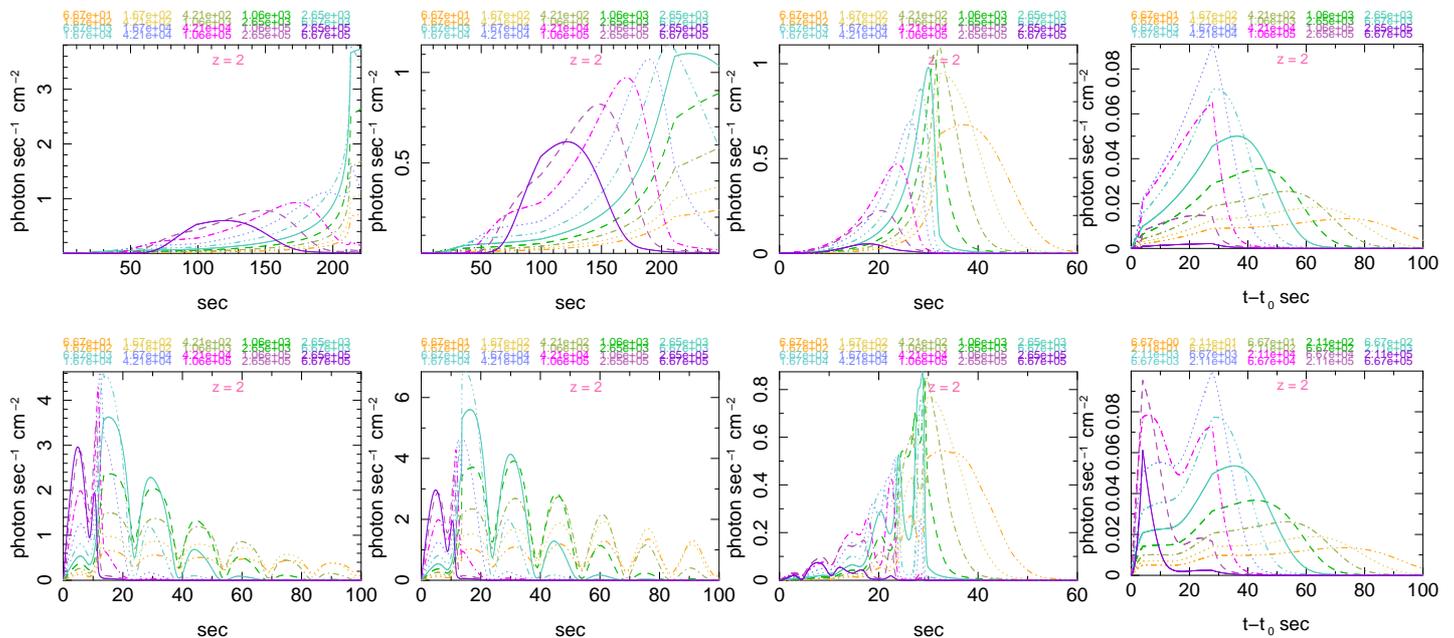

\begin{center}
\begin{tabular}{llll}
\hspace{-1.5cm} \includegraphics[width=4cm,angle=-90]{\bdir/spec-m1p-102-lin-lc-mag0-0.eps} & 
\hspace{-0.8cm} \includegraphics[width=4cm,angle=-90]{\bdir/spec-m1p-122-lin-lc-mag0-0.eps} &
\hspace{-0.8cm} \includegraphics[width=4cm,angle=-90]{\bdir/spec-m7-122-lin-lc-mag0-0.eps} & 
\hspace{-0.8cm} \includegraphics[width=4cm,angle=-90]{\bdir/spec-m2p-1222-lin-lc-mag0-0.eps} \\
\hspace{-1.5cm} \includegraphics[width=4cm,angle=-90]{\bdir/spec-m1p-122-lin-lc-mag600-02.eps} &
\hspace{-0.8cm} \includegraphics[width=4cm,angle=-90]{\bdir/spec-m1p-102-lin-lc-mag600-02.eps} &
\hspace{-0.8cm} \includegraphics[width=4cm,angle=-90]{\bdir/spec-m7-122-lin-lc-mag100-05.eps} &
\hspace{-0.8cm} \includegraphics[width=4cm,angle=-90]{\bdir/spec-m2p-1222-lin-lc-mag20-01.eps}
\end{tabular}
\caption{\small Top row: GRB simulations without an external magnetic field. Each simulation includes 
3 regimes. The lags in the first two models from left is too large to be consistent with 
observations. Bottom row: Simulations with a precessing external magnetic field. From left to 
right: 1) $|B| = 100$~kG, $f=0.2$ Hz 2) $|B| = 12$~kG, $f=0.2$ Hz, and middle regime dynamical, 
3) $|B| = 100$~kG, $f=0.2$ Hz, and middle regime steady state, 4) $|B| = 2.5$~kG, $f=0.1$ Hz. 
Other parameters are the same as simulations in the top row. In all plots of light curves 
energy bands are written on the top of the plot and the color of their font corresponds to 
the color of light curves of the band \label{fig:lcmag}}
\end{center}
\end{figure*}

\begin{table}
\caption{Parameter set of the phenomenological shock model \label{tab:param}}
\begin{center}
\begin{tabular}{p{7mm}p{4mm}p{7mm}p{11mm}p{7mm}p{7mm}p{10mm}p{10mm}p{10mm}p{5mm}p{5mm}p{5mm}p{5mm}p{5mm}p{7mm}}
\hline
model & $r_0$ (cm) & $\frac{\Delta r'_0}{r'_0} $ & $p$ & $\alpha_p$& $\gamma_{cut}$ & $\kappa$ & $ \gamma'_0$ & $ \tau$ & $ \delta$ & $\epsilon_B$ & $\alpha_B$ \\
\hline 
$\epsilon_e$ & $\alpha_e$ & $N'_0$ (cm$^{-3}$) & $n'\Delta r'(r'_0)$ (cm$^{-2}$) & 
$\Gamma_f$ & $|B|$ (Gauss) & $f$ (Hz) & $\alpha_x$ & phase (rad.) & $(\frac{r'}{r'_0})_{max}$ & & \\
\hline
\end{tabular}
\end{center}
\end{table}
Fig. \ref{fig:lcmag} show few examples of simulated bursts without and with an external 
magnetic fields. It is clear that in presence of an external field the bursts are usually 
brighter, harder, and lags between the light curves of the various energy bands are in 
general smaller and more consistent with observations. Nonetheless, some simulated bursts 
have small lags even in absence of an external field. Therefore, its presence is not a 
necessary condition for the formation of a GRB. The light curves of these examples show 
that despite the presence of a precessing field - similar to what is seen in pulsars and 
magnetars - there is barely any signature of oscillation in these light curves, specially 
in high energy bands. Only when the duration of the burst is a few times longer than 
oscillation period, the precession of the field creates a detectable periodic component in 
the light curves, see Fig. \ref{fig:shortgrab}. We also note that oscillations are more 
visible in soft X-ray and UV/optical bands. However, only in the case of the detection of a 
well separated precursor low energy data are available during the prompt emission. 
Moreover, the necessity of binning of optical data to reduce the noise can smear fast 
oscillations, see Fig. \ref{fig:shortgrab}. Therefore such oscillation can be hardly 
observed in real GRBs. On the other hand, in many burst e.g. GRB 070129 semi-periodic 
variations of the early time X-ray light curves, similar to two of the examples in 
Fig. \ref{fig:lcmag}, have been observed for a few hundreds of seconds. 

\begin{figure*}
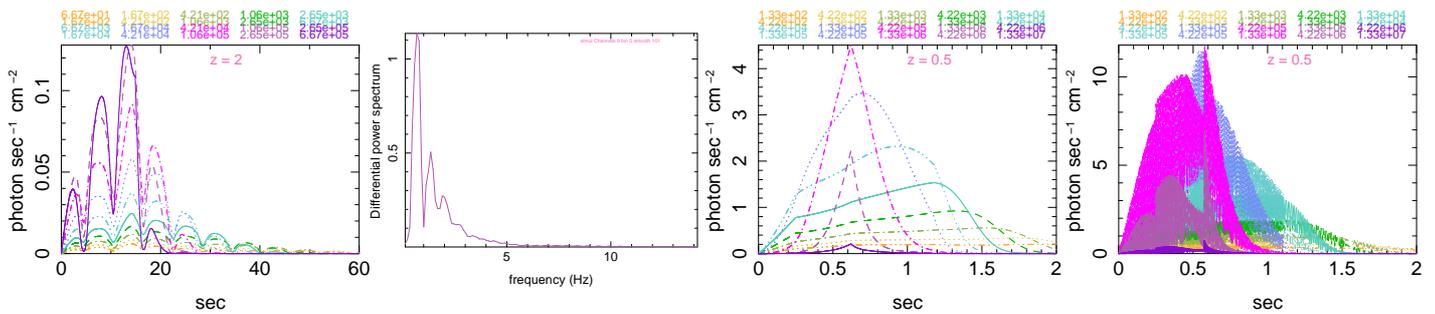

\begin{center}
\begin{tabular}{llll}
\hspace{-1.cm}\includegraphics[width=4cm,angle=-90]{\bdir/spec-m5-1022-log-lc-mag200-05.eps} &
\hspace{-0.5cm}\includegraphics[width=3.7cm,angle=-90]{\bdir/spec-m5-1022-log-lc-mag200-05_1_10_1simul.ps} &
\hspace{-0.5cm}\includegraphics[width=4cm,angle=-90]{\bdir/spec-m4-1022-lin-lc-mag0-0.eps} &
\hspace{-0.5cm}\includegraphics[width=4cm,angle=-90]{\bdir/spec-m4-1022-lin-lc-mag200-500.eps} 
\end{tabular}
\caption{\small From left to right: 1) Simulation of a long burst with a precessing magnetic 
field and 2) its the Power Distribution Function (PDF) of the total light curve. Oscillatory 
component in high energy bands is visible by eye and in the PDF. 3,4) Simulations of a short 
burst without and with a fast precessing $f=500$ Hz external magnetic field. The 
comparison of two plots show that due to nonlinearity of the dynamics the magnetic field 
induces substructures at time scales much longer than the precession period. 
\label{fig:shortgrb}}
\end{center}
\end{figure*}

\begin{figure*}
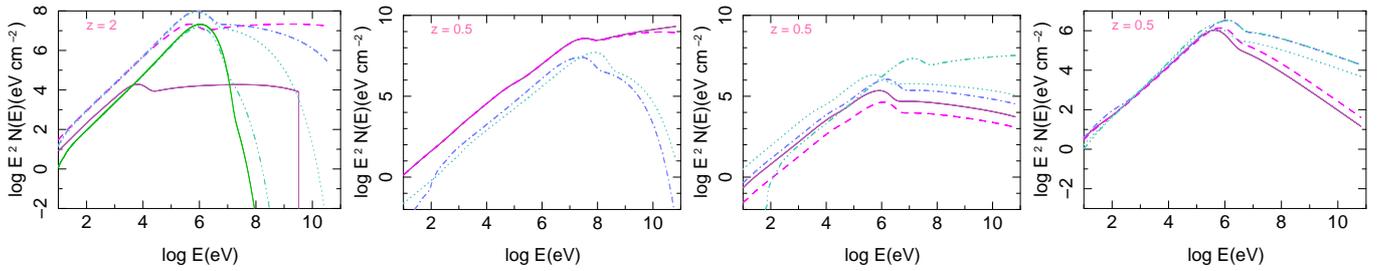

\begin{center}
\begin{tabular}{llll}
\hspace{-1cm}\includegraphics[width=3.5cm,angle=-90]{\bdir/spec-m1p-122-mag60-02-expocut-highener-4.ps} & 
\hspace{-0.3cm}\includegraphics[width=3.5cm,angle=-90]{\bdir/m3-322-variouscut-4.ps} & 
\hspace{-0.3cm}\includegraphics[width=3.5cm,angle=-90]{\bdir/m3-322-variousepsilon-4.ps} & 
\hspace{-0.3cm}\includegraphics[width=3.5cm,angle=-90]{\bdir/m3-322-plcut-tot.ps}
\end{tabular}
\caption{\small From left to right: 1) Electron distribution: power-law with exponential 
cutoff, $p = 2.5$ and $\omega_{cut}/\omega_m =$ 0.5 (full line), 1 (dash-3 dots), 10 (dot), 
100 (dash-dot); $p = 1.9$ and $\omega_{cut}/\omega_m = 1000$ (dash). 
The external magnetic field in these simulations is $10$ kG. The low amplitude full line has 
$p = 1.9$ and $\omega_{cut}/\omega_m = 1000$ but no external magnetic field. 2) Electron 
distribution: power-law with exponential cutoff for electrons, $|B_{ext}| = 70$ kGauss: 
$\omega_{cut}/\omega_m = 1000$, $p = 1.5$ (full line); $\omega_{cut}/\omega_m = 100$, $p = 1.5$ 
(dash); $\omega_{cut}/\omega_m = 3$, $p = 2$ (dot-dash); $\omega_{cut}/\omega_m = 3$, $p = 2.5$ 
(dot). 3) Electron distribution: power-law with exponential cutoff, $|B_{ext}| = 100$ kGauss: 
$\omega_{cut}/\omega_m = 1000$, $p = 2.5$, $\epsilon_e = 0.002$, $\Gamma = 500$ (full line); 
$\omega_{cut}/\omega_m$ and $p$ as previous case and $\epsilon_e = 0.02$, $\Gamma = 50$ 
(dash); $n'_0 = 5 \times 10^{15}$ cm$^{-3}$ and other parameters as the previous case 
(dot-dash); varying $p$ with index $-0.2,~0,~0.5$, initial $p = 2.5$ and 
$\epsilon_e = 0.002$ (dot); initial $p = 1.8$, the same indices as previous, and 
$\omega_{cut}/\omega_m = 0.5,~1000,~100$ (dash-3 dots). 4) Electron distribution: broken 
power-law a broken slope at $\omega_{cut}/\omega_m = 3$ and $p_1 = 2.5$, $p_2 = 4$, 
$|B_{ext}| = 17$ kGauss (full line); $p_1 = 2.1$, $p_2 = 4$ and same $|B_{ext}|$ as previous 
(dash); $p_1 = 2.1$, $p_2 = 3$, $|B_{ext}| = 26$ kGauss (dot-dash); same slope and 
$|B_{ext}| = 35$ kGauss (dot); same slope and $|B_{ext}| = 70$ kGauss (dash-3 dots). 
\label{fig:spect}}
\end{center}
\end{figure*}

Our model can be 
applied to both long and short bursts. An example of simulation of short bursts is shown in 
Fig. \ref{fig:shortgrb}. The lags in this example is very short - consistent with observed 
zero lags - both in presence and absence of an external magnetic field. It also shows that 
very fast precession of the field can be confused with the shot noise. A remarkable 
properties of simulation with an external magnetic field is the presence of non-periodic 
substructures is the light curves from nonlinearity and fast variation of physical quantities. 
Fig. \ref{fig:spect} shows some examples of spectra obtained for simulated bursts. They have 
a variety of behaviour at high energies. Notably, when the cutoff energy is high and the 
spectrum is flatter than what is possible for a simple power-law i.e. $p \leqslant 2$, the 
slope of the fluence at very high energies is positive, i.e. flux increases. Evidently, even 
in this case at very high energies the spectrum bends and the slope becomes negative. An 
example of such cases is the dash-dot spectrum in the first plot of Fig. \ref{fig:spect}.

\subsection{Compton scattering, delayed high energy component, and other issues}
We have also simulated the effect of the inverse Compton scattering of photons by high energy 
electrons, but due to the limited length of this proceeding we do not show them here. As we 
mentioned in Sec. \ref{sec:shock} it is crucial to consider a realistic distribution of 
accelerated electrons along the photons path, otherwise we overestimate the scattering rate 
and obtain a spectrum that is not consistent with observations. Nonetheless, the light curves 
of IC scattered photons are more extended in time. Thus, apriori they should explain the 
detected delayed emission at high energies. However their flux is much smaller then what is 
observed. 

On the other hand, the similarity of spectra shown in Fig. \ref{fig:spect} which include only 
the synchrotron emission to observations is the evidence that the origin of delayed high 
energy emission is the same as lower energies. In~\cite{hourigrb2} we have given detailed 
arguments that the reason for the delay of high energy emission is that the most energetic 
electrons are trapped in the EES and follow its propagation. In fact this is a 
self-organizing 
processes: energetic electrons can follow the propagation of EES. In this way they stay in 
the region where electric field is strong and do not get a lag that brings them to the region 
where magnetic field is strong. But, because they stay longer in the high electric field 
region, they are accelerated more and can better follow the propagation of EES. Simple 
calculations show that for nominal value of parameters the delay can be tens of seconds 
consistent with observations. We leave a quantitative study of this process for a future work.
The issues of the slope at low energies and efficiency are also discussed 
in~\cite{hourigrb2} and we show that they can be explained in the context of SSC model.
\small
\bibliographystyle{apalike}

\begin{thebibliography}{10}
\bibitem [Zhao, \etal 2006]{snhistory} Zhao FY, Strom RG, Jiang SY (2006), The Guest Star of AD185 Must Have Been a Supernova", \Journal{\CJA} {6}{(5)}{635–640}.
\bibitem [da Silva 1993]{sntyperev} da Silva, L.A.L. (1993), The Classification of Supernovae, \Journal {\ASS}{202}{(2)}{215–236}.
\bibitem [Klebesadel \etal 1973]{grbdetect} Klebesadel R.W., Strong I.B., and Olson R.A. (1973), Observations of Gamma-Ray Bursts of Cosmic Origin , \Journal {\APJ}{182}{}{L85}.
\bibitem [Bonnell 1995]{grbhist} Bonnell J. (1995), http://apod.nasa.gov/htmltest/jbonnell/www/grbhist.html.
\bibitem [Paczynski 1986]{cosminsn} Paczynski, B. (1986), Gamma-ray bursters at cosmological distances, \Journal {\APJL}{308}{(2)}{L43-L46}.
\bibitem [Goodman 1986]{fireball1} Goodman, J. (1986), Are gamma-ray bursts optically thick? \Journal{\APJL}{308}{(2)}{L47}.
\bibitem [Piran 1999]{piranrev} Piran, T (1999), Gamma-Ray Bursts and the Fireball Model, \Journal{\PRE}{314}{}{575} [astro-ph/9810256].
\bibitem [Rees \& M\'esz\'aros 1994]{intext} Rees M.J. \& M\'esz\'aros P. (1994), Unsteady Outflow Models for Cosmological Gamma-Ray Bursts, \Journal{\APJ}{430}{}{L93} [astro-ph/9404038].
\bibitem [Ziaeepour 2009a]{hourigrb} Ziaeepour H. (2009a), Gamma Ray Bursts Cook Book I: Formulation, \Journal {\MRA}{397}{2009}{361} [arXiv:0812.3277].
\bibitem [Ziaeepour 2009b]{hourigrb1} Ziaeepour H. (2009b), Gamma Ray Bursts Cook Book II: Simulation, \Journal {\MRA}{397}{2009}{386} [arXiv:0812.3279].
\bibitem [Ziaeepour \& Gardner 2011]{hourigrb2} Ziaeepour H. \& Gardner B. (2011), Broad band simulation of Gamma Ray Bursts (GRB) prompt emission in presence of an external magnetic field, \Journal{\JCA}{12}{}{001}.
\bibitem [Fenimore \& Ramirez-Ruiz 1999]{shelldecel} Fenimore E.E. \& Ramirez-Ruiz E. (1999), Gamma-Ray Bursts as internal shocks caused by deceleration, [astro-ph/9909299].
\bibitem [Umeda \etal 2005]{precursdecel} Umeda, H., Tominaga, N., Maeda, K., \& Nomoto, K. (2005), Precursors and main bursts of Gamma-Ray Bursts in a hypernova scenario, \Journal {\APJ}{633}{2005}{L17}, astro-ph/0509750.
\bibitem [Medvedev \& Loeb 1999]{bgrmagfield} Medvedev M. \& Loeb A. (1999), Generation of Magnetic Fields in the Relativistic Shock of Gamma-Ray-Burst Sources, \Journal{\APJ}{526}{}{697} [astro-ph/9904363].
\bibitem [Lyutikov \& Blandford 2003]{poytingflow} Lyutikov M. \& Blandford R.D. (2003), Gamma Ray Bursts as Electromagnetic Outflows, [astro-ph/0312347].
\bibitem [Sari \etal 1996]{emission0} Sari R., Narayan R., \& Piran T. (1996), Cooling Time Scales and Temporal Structure of Gamma-Ray Bursts, \Journal{\APJ}{473}{}{204} [astro-ph/9605005].
\bibitem [Sakamoto \etal 2008]{batcat} Sakamoto T., \etal (2008), The First Swift BAT Gamma-Ray Burst Catalog, \Journal{\APJS}{175}{}{179} [arXiv:0707.4626].
\bibitem [Spitkovsky 2008]{fermiaccspec} Spitkovsky A. (2008), Particle acceleration in relativistic collisionless shocks: Fermi process at last?, \Journal{\APJ}{682}{}{5} [arXiv:0802.3216].
\bibitem [Murphy \etal 2010]{fermiaccspec1} Murphy G.C., Dieckmann M.E., Drury L. O'C. (2010), Multidimensional simulations of magnetic field amplification and electron acceleration to near-energy equipartition with ions by a mildly relativistic quasi-parallel plasma collision, IEEE Transactions on Plasma science, 38, 2985 [arXiv:1011.4406].
\end{thebibliography}
 
\end{document}